\documentclass[12pt]{article}
\usepackage{openwork}

\hypersetup{hidelinks}

\title{Towards a More Realistic VR Experience: Merging Haptic Gloves with Precision Gloves \thanks{Presented as Abstract P3-24 at the 10th International Symposium on Biomedical Engineering (ISBE2025) and the International Workshop on Nanodevice Technologies 2025 (IWNT2025), October 30--31, 2025, Higashihiroshima, Japan.}}
\author[1]{P.Bottoni}
\author[1]{S.Cifani}
\author[2]{K.Kanev}
\author[3]{D.Moraru}
\author[3]{A.Nakamura}
\author[1]{M.R.Marini}
\affil[1]{Department of Computer Science, Sapienza University of Rome, Rome, Italy}
\affil[2]{Faculty of Business and Information Technology, Ontario Tech University, Canada}
\affil[3]{Research Institute of Electronics, Shizuoka University, Japan}
\date{}

\begin{document}

\maketitle

\begin{center}
    \vspace{-0.7em} 
    \small (bottoni, cifani, marini)@di.uniroma1.it;\\
    (kanev, moraru.daniel, nakamura.atsushi)@shizuoka.ac.jp
    \vspace{0.5em}
\end{center}

\textbf{Introduction:} Virtual reality (VR) glove technology is increasingly important for professional training \cite{boutin2024smart}, industrial applications \cite{civelek2022virtual}, and teleoperation in hazardous environments \cite{xu2025immersive}, since it enables more natural and immersive interactions than controllers. However, current solutions face a trade-off: high-precision gloves lack haptic feedback, while haptic gloves suffer from poor accuracy. Existing studies have mainly focused on developing new glove prototypes or optimizing only one type of glove, without addressing the integration of both features. Our work presents a novel hybrid approach that combines a high-precision glove with a haptic glove, creating a system that delivers both precision and haptics.

\begin{figure}[h]
  \centering
  \begin{minipage}[t]{7.5cm} 
    \centering
    \includegraphics[width=10cm]{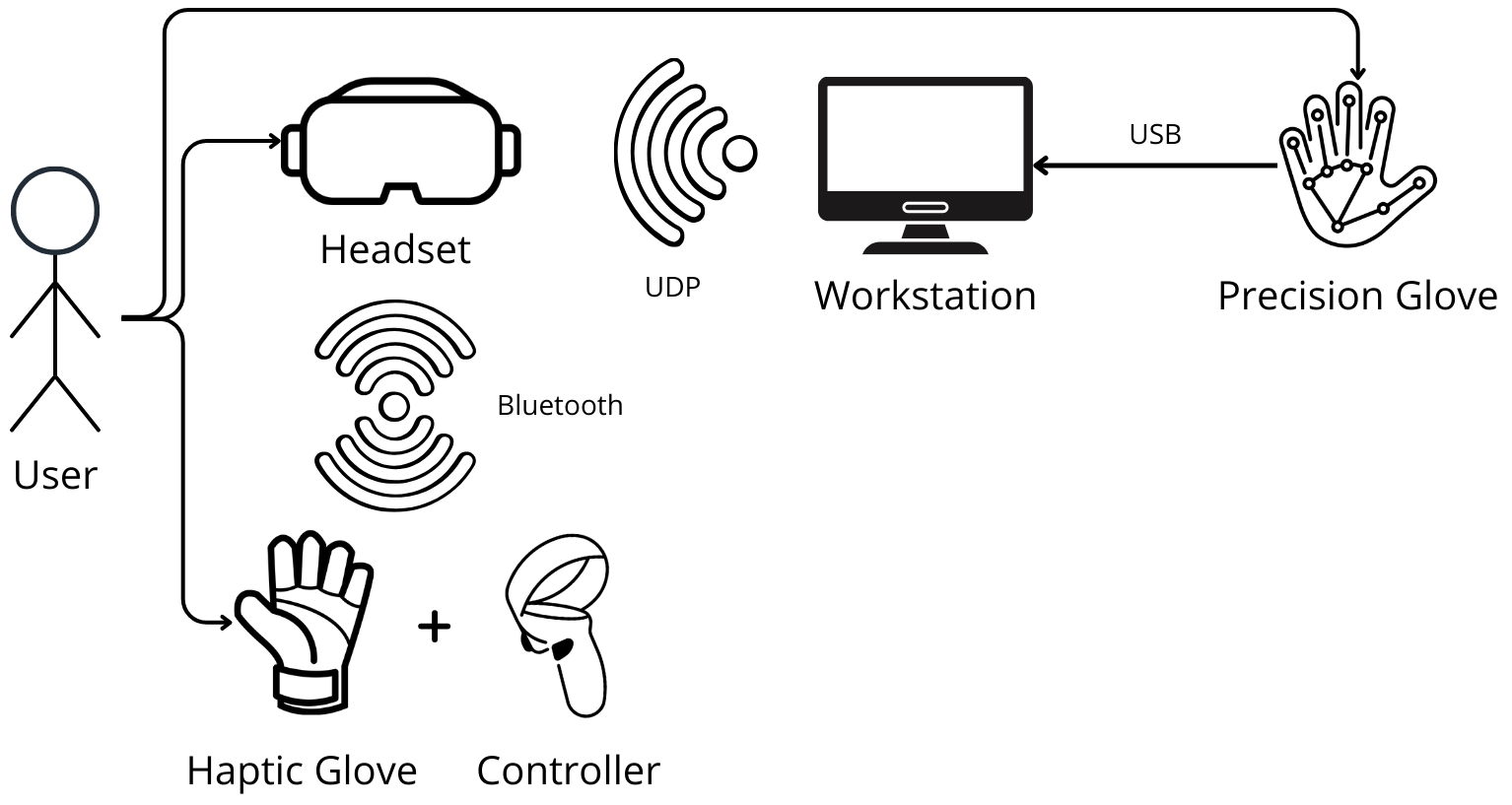} 
    \caption{Hybrid system overview}
    \label{fig:overview}
  \end{minipage}
  \hfill 
  \begin{minipage}[t]{6cm}
    \centering
    \includegraphics[width=6.5cm]{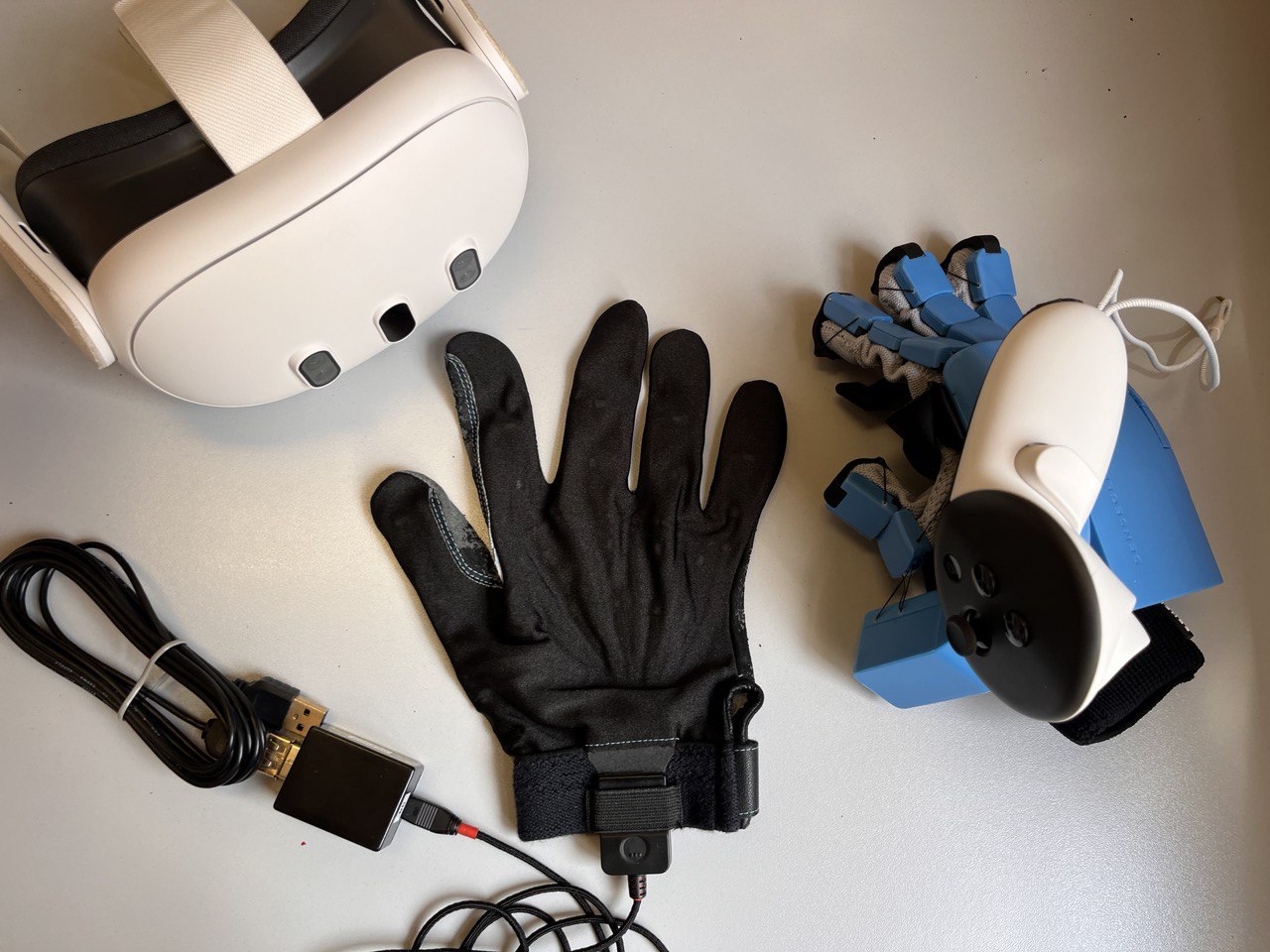} 
    \caption{Hybrid system hardware}
    \label{fig:hardware}
  \end{minipage}
\end{figure}

\textbf{Materials and Method:} To address the limitations of standalone gloves, we have developed a hybrid system (see Fig. \ref{fig:overview}). The experimental setup (see Fig. \ref{fig:hardware}) consists of a VR workstation (PC), a Meta Quest 3 headset, and its controller for spatial tracking. Finger flexion is captured with a high-precision Data Glove by Yamaha \cite{gelsomini2021specialized}, while a haptic glove Nova 1 by SenseGlove delivers force feedback. In this configuration, the user wears the Yamaha glove underneath the Nova 1, which has the spatial tracker mounted on it.
Our primary technical contribution is the method for data integration. Finger-bending data from the precision glove is continuously acquired on the PC and streamed via UDP to the Unity application running on the VR headset. There, we inject the high-fidelity data directly into code adapted from the SenseGlove Software Development Kit library incorporated into Unity. This novel approach bypasses the haptic glove's native, imprecise input while retaining its immersive force feedback. The system has been evaluated using a Unity-based industrial component assembly task, with user testing conducted via the System Usability Scale (SUS) \cite{lewis2018system} and Raw NASA Task Load Index (NASA-TLX) to assess usability and workload.

\textbf{Results and Discussion:} Our user study involved 15 participants (10 males, 5 females) from different technical (e.g., Computer Science) and non-technical (e.g., Sociology, Food Science, Communication) backgrounds. For analysis, participants were categorized by their VR experience: inexperienced (never used VR or used it only once; n=8), average (occasional users; n=4), and experienced (frequent users; n=3).

\begin{table}[h]
    \centering
    \caption{SUS and Raw NASA-TLX scores by user type}
    \label{tab:results}
    \begin{tabular}{lcc}
        \toprule
        \textbf{User Type} & \textbf{SUS Score} & \textbf{Raw NASA-TLX Score} \\
        \midrule
        Inexperienced VR Users & 74.28\% & 25.39\% \\
        Average VR Users & 55.83\% & 32.71\% \\
        Experienced VR Users & 67\% & 25.18\% \\
        \midrule
        \textbf{All users} & \textbf{65.70\%} & \textbf{27.76\%} \\
        \bottomrule
    \end{tabular}
\end{table}

The overall results (see Table \ref{tab:results}) were promising, indicating the system was perceived as easy to use and imposed a low cognitive load, with a mean SUS score of 65.70\% and a mean Raw NASA-TLX score of 27.76\%. A detailed analysis revealed a clear pattern based on user experience. Inexperienced users were highly enthusiastic (likely due to a lack of a comparative baseline) and quickly overcame their initial intimidation, learning the system with ease. This was reflected in their high SUS score of 74.28\% and low Raw NASA-TLX score of 25.39\%. Average users, in contrast, provided the lowest ratings (55.83\%) and reported the highest perceived workload (32.71\%). This suggests that, while their experience gave them a point of comparison, they lacked the advanced expertise to navigate the system's limitations effectively. Experienced users leveraged their advanced knowledge to manage the system's complexities, rating it more favorably than the average group (67\%) and reporting a low workload (25.18\%). Consequently, their qualitative feedback consisted of specific technical suggestions rather than general usability comments. Finally, a notable finding across inexperienced and average users was that the dual-glove setup was not perceived as bulky or cumbersome.

\textbf{Conclusions:} This study demonstrated that our hybrid approach is a promising solution to the critical limitations of standalone haptic gloves. Prior to this work, the use of force-feedback gloves alone resulted in jittery virtual hand mapping and a stressful user experience, making them unsuitable for precision tasks. By integrating a high-precision data glove, we achieved a significant enhancement in finger tracking fluidity and accuracy. While this system serves as a successful proof-of-concept, future work will focus on refining the calibration function, minimizing the already low latency, and developing a fully wireless solution. In conclusion, our work transforms a previously unusable technology into a promising tool for VR applications that demand both high fidelity and immersive haptic feedback.


\newpage
\printbibliography[title={{\fontsize{12}{16}\selectfont References:}}]

\textbf{Biography:} P. Bottoni and R.M.Marini are, respectively, Full Professor and Researcher of Computer Science at Sapienza University of Rome. S. Cifani is a Ph.D. Student in AI at Sapienza. K. Kanev is Adjunct Professor at Ontario Tech University. D. Moraru and A. Nakamura are Professors at Shizuoka University.

\textbf{Acknowledgment:} This study was supported by the Research Center for Biomedical Engineering.

\end{document}